**DC Leakage behavior and Conduction Mechanism in $(BiFeO_3)_m(SrTiO_3)_m$ Superlattices**

R. Ranjith and W. Prellier[1]

*Laboratoire CRISMAT, CNRS UMR 6508, ENSICAEN*
*6 Bd Maréchal Juin, 14050 Caen Cedex, France.*

Jun Wei Cheah and Junling Wang

*School of Materials Science and Engineering*
*Nanyang Technological University, Singapore-639798.*

Tom Wu

*Division of Physics and Applied Physics,*
*School of Physical and Mathematical Sciences*
*Nanyang Technological University, Singapore-637616.*

Leakage current behavior of $(BiFeO_3)_m(SrTiO_3)_m$ superlattice structures was studied and analyzed at different temperatures (303-473K) in the light of various models. While bulk limited Poole-Frenkel emission was observed to dominate the leakage current in the temperature range of 303 – 383 K, the space charge limited conduction was observed up to 473 K. With a Poole-Frenkel emission type of conduction, the activation energy range of ~ 0.06 – 0.25eV was calculated. The physical parameters, calculated from the analysis correlates with the intrinsic properties. Such analysis of leakage current facilitates interface engineering of heterostructures for device applications.

---

[1] prellier@ensicaen.fr

Leakage current analysis in a ferroelectric thin film has been an important study for the use of ferroelectric materials in functional devices.[1] There has been a flurry of studies in the literature in connection with understanding of the leakage mechanisms in ferroelectric thin films.[2,3,4,5] The major leakage mechanisms in insulating thin films are broadly classified into three categories: the electrode-material interfacial barrier limited Schottky emission, the modified Schottky conduction,[6] the space charge limited bulk conduction, and the Poole Frenkel (PF) emission arising from the bulk.[1,2] The Schottky emission is known as an interfacial phenomenon that arises due to the barrier formation at the material electrode interface.[1,2] The Poole-Frenkel and space charge limited (SCL) conduction arises from the bulk of the material under study.[3-5]

$BiFeO_3$ (BFO) has been extensively studied in recent years due to its room temperature multiferroic behavior.[7,8] It is one of the rarely existing natural room temperature multiferroic materials with a magnetic Neel Temperature of 643K and a high ferroelectric transition temperature of 1103K.[7] In spite of its superior ferroelectric properties it often shows a high dc leakage current.[9] The high DC leakage of $BiFeO_3$ could be attributed to the presence of $Fe^{3+}$ cations with a non $d^o$ electronic configuration, which hinders the use of this material in ferroelectric memory devices.[9] Various methods have been employed to reduce the leakage and to understand the underlying mechanisms.[10,11,12,13] Doping aliovalent impurities on both A and B sites of BFO has been an effective approach to reduce the leakage current, in both bulk and thin films.[10,11]

Whereas, the leakage mechanism was unclear, recently a Poole-Frenkel emission was considered as the leakage mechanism in a BFO thin film epitaxially grown on $SrRuO_3$ bottom electrode with $SrRuO_3$ and Pt as top electrodes[9] and in a thin film of BFO and $PbTiO_3$ solid solution.[11] The Poole-Frenkel emission as the dominant mechanism of leakage current in ferroelectric thin films is not a new phenomena and has been observed in Lead Zirconate thin films[3,4]. Among the various approaches to reduce the leakage current in BFO thin films, recently a superlattice with a combination of BFO and SrTiO$_3$ (STO) has been proven to be a useful approach to improve the leakage behavior of BFO.[12] However, the leakage mechanism of the BFO-STO superlattice structures remained ambiguous. Furthermore, understanding the leakage mechanisms in BFO related materials is crucial for interfacial engineering for device applications.

In this work, superlattice structures of $(BiFeO_3)_m(SrTiO_3)_m$ were fabricated through the pulsed laser deposition technique. Superlattices with different periodicities were fabricated and the temperature dependence of the DC leakage behavior was studied. The leakage mechanism was analyzed based on the different models specific to ferroelectric and insulating thin films. Finally, the physical parameters of the system under study were extracted from the fitting.

Thin films of BFO and STO forming superlattices were synthesized by repeating several times a bilayer consisting of m unit cells of BFO and STO, where m takes integer values from 1 to 10. The total thickness of the superlattice was kept constant at 2400Å. The details of the fabrication of thin films and device processing can be found elsewhere.[12] The fabricated heterostructures were characterized in the metal-insulator-metal (MIM) configuration to study their leakage current (Current – Voltage (I-V)) behavior. The room temperature ferroelectric polarization hysteresis (P-E loops) has been reported elsewhere.[12] The leakage current was measured with delay time corresponding to a minimum relaxation current, which is commonly known to overlap with the leakage current.[4]

Figure 1 shows the I-V curves of a superlattice structure with m=5 (Λ=40Å) measured in the temperature range of 303 K - 473 K. The inset shows the symmetric nature of the I-V curve on reversing the bias voltage. Despite of an asymmetric structural configuration the symmetric nature on reversing the bias shows that the observed leakage behavior is not dominated by the asymmetric electrode material interface. Figure 2 shows the modified

Schottky plot with ln(J/E) vs. $E^{1/2}$ for m = 5 and 7, and the inset shows the activation plot ln(J/ET$^{3/2}$) vs. 1000/T derived from the modified Schottky equation as given in equation (1).[6] A normal Schottky plot is limited to a system with an electronic mean free path larger than the inter electrode spacing. Hence, a modified Schottky plot with a factor of applied field included in the pre-exponential term of the normal Schottky plot was used considering the system under study (BFO-STO).[6] The applied electric field dependence of the barrier which, forms at the cathode due to the difference in the Fermi energies of the metal electrode and the insulator is taken into account in the modified equation.[6] In-spite of the correction, the modified Schottky equation does not clearly distinguishes between the electrode limited and the bulk phenomena.[14] The current density given by the modified Schottky equation [6,14] is given by

$$J = \alpha\mu E T^{3/2} \left(\frac{m^*}{m_o}\right)^{3/2} \exp\left\{-\frac{\Phi}{kT} + \frac{1}{kT}\left(\frac{e^3 V}{4\pi\varepsilon_o K d}\right)^{1/2}\right\} \quad \ldots \quad (1)$$

where, $\alpha = 3 \times 10^{-4}$ As/cm$^3$K$^{3/2}$, e – electronic charge, µ- the electron mobility, $m^*$ and $m_o$ the effective and free electron mass respectively, $\varepsilon_o$- permittivity of free space, K – high frequency dielectric constant, k – Boltzmann constant, $\Phi$ – barrier height.[6] The fitting of I-V curve with the current density equation given above, renders physical parameters like, the refractive index and the high frequency dielectric constant. However, the validity of Schottky emission should be determined by the physical parameters derived from the fitting, which is a characteristic of the material under study. The modified equation does not differentiate between the interface and the bulk effects occuring in the sample. The derived high frequency dielectric constant and the refractive index have to be consistent with the intrinsic material properties. The refractive index of BFO is known to be around 2.5 and the high frequency dielectric constant is expected to be around 6.25.[9] Although the low temperature linear fitting is reasonably good,(as seen in Figure 2) the refractive index and the corresponding high frequency dielectric constant derived from the Schottky equation were in the range of ~ 0.78 – 1.2 and ~ 0.62 – 1.4, respectively.

Besides the fact that the observed values were one order of magnitude smaller, they were physically unrealistic in terms of refractive index and high frequency dielectric constant. The barrier height (Φ) obtained from the activation plot (Inset of Figure 2) was found to be in the range of ~ 0.014eV – 0.063eV at different applied electric field. Hence, the modified Schottky emission could be ruled out as the dominant leakage mechanism in the BFO-STO superlattice structures.

The other mechanism attributed to the leakage current in insulating thin films is the Poole-Frenkel emission. It is a bulk limited process in which the emission of charge carriers trapped in the defect centers, contributes to the conduction process. The trap centers could be distributed in the forbidden region between the valence band and the conduction band of the material. The carriers in the traps could be activated either thermally or electrically. Under an electric field, at a given temperature, the ionization of traps induces the emission of charge carriers and gives rise to conduction. The conductivity due to the bulk limited PF emission is given by

$$\sigma = \sigma_o \exp\left\{\left(\frac{-E_t}{kT}\right) + \beta_{PF}\sqrt{E}\right\} \qquad \ldots \qquad (2)$$

Where, $\sigma_o$ is the conductivity at zero field and $\beta_{PF} = \frac{1}{kT}\left(\frac{e^3}{\pi\varepsilon_o K}\right)^{1/2}$ and $E_t$ the trap ionization energy.[4,9] Figure 3 shows the Poole-Frenkel fitting of the BFO-STO superlattice structures for m = 5 recorded at different temperatures. The fitting is reasonably good for a wide range of voltage and temperature. The validity of the mechanism could be verified by the magnitudes of the characteristic physical entities derived from the curves. The range of high frequency dielectric constant and the refractive index derived from the PF type of conduction in the temperature range of 300 – 383K are 5.7 - 7.1 (6.5-BFO)[15] and 2.3 - 2.6 (2.5-BFO, 2.2-2.6 – STO)[16] respectively. The observed values correlate well with the intrinsic material properties of BFO and STO, available in the literature and with reported values for the single layer BFO thin films.[15] The inset of Figure 3 shows the thermal activation of the conductivity and was found to be in the range of ~ 0.06 to 0.25eV. The energy range observed is lower in comparison to the energy obtained for a single layer BFO (~ 0.65-0.8eV).[9] In the case of single layer BFO thin films, the defects are expected to originate from the oxygen vacancy formed due to the mixed oxidation state of $Fe^{2+}$ and $Fe^{3+}$ cations.[9] In the case of superlattice structures, in addition to the oxygen vacancies, the defects could arise from the high trap densities due to the strain fields at the interface, defects like misfit dislocations due to strain relaxation, and the oxygen vacancies at the interface. The energy band gaps of BFO and STO are known to be ~ 2.5eV[15] and ~3.5 eV[16], respectively. Hence, a distribution of shallow traps with low activation energies could be expected at the interface between BFO and STO. Interface of the BFO and STO is expected to be the source of strain field and defects where, the charge carriers could be trapped and effectively dominate the leakage behavior.

On observing the I-V curves and the PF fitting in the range of 413 – 473 K, we conclude that at higher temperatures (> 383K) the transport is not dominated by the PF emission. In the case of insulating thin films on application of higher electric fields, the conduction is dominated by the space charge (detrapped carriers and/or free charges from the electrode) conduction.[4] In the case of space charge limited conduction the current density obeys the power law dependence ($J \propto V^n$).[17]

Beyond certain applied electric field, the charges are detrapped into the conduction band giving rise to a sudden increase in the leakage current density. The region corresponds to the n = 2 limit, where, the current arises due to the excess space charge that are injected into the conduction band and is referred as space charge limited conduction.[17] As the voltage increases beyond the trap free level, the current injection is mainly due to the excess charges present in the conduction band and the current density at the trap free limit is governed by the Child's law, described as:[17]

$$J = \frac{9\mu\varepsilon_o \varepsilon V^2}{8d^3} \qquad \ldots \qquad (3)$$

Where, µ is the mobility of the charge carriers, V the applied voltage, and d the thickness of the material, in our case the total thickness of the superlattice. Figure 4 shows the logJ - logV plot of the BFO-STO superlattice with m = 5. The data exhibited different regions with n values varying from 1.15 ± 0.03 to 4.25 ± 0.03 at different magnitudes of applied field. At 473K, a narrow trap free region with n = 4.25 ± 0.03 followed by a space charge limited region, with n = 2.01 ± 0.03 at high field, was observed. Thus, at temperatures above 383 K and at higher applied field the leakage was observed to be dominated by space charge limited conduction with n = 2.01 ± 0.03.[17]

In summary, $(BiFeO_3)_m(SrTiO_3)_m$ superlattice structures fabricated by pulsed laser ablation were used to study the leakage current behavior. The temperature-dependence of the DC leakage current was investigated at different applied voltages. The conduction mechanism of the leakage current was analyzed in the light of existing models for the transport in insulating thin film capacitors. It was observed that the bulk limited Poole-Frenkel emission dominates the conduction in the temperature range of 303K – 383K and above which, up to 473K, the leakage current conduction was observed to be dominated by space charge limited conduction at higher applied field. The emission of charge carriers from the traps distributed in the forbidden region dominates the leakage current conduction due to the high density of interfaces in a superlattice structure. The observed results will be useful for reducing the undesired leakage through interfacial engineering and will benefit device applications.

This work was carried out in the frame of the Merlion project (n°2.04.07) supported by the Ministere des Affaires Etrangeres et Européennes and the Nanyang Technological University; the STREP MaCoMuFi (NMP3-CT-2006-033221) supported by the European Commission and by the CNRS, France. Partial support from the ANR (NT05-1-45177, NT05-3-41793) is also acknowledged. The authors would also like to acknowledge Dr. L. Mechin, and Mr. C. Fur.
**Reference:**

[1] M.Dawber, K.M. Rabe and J.F. Scott, Rev. Mod. Phys., **77**, 1083 (2005)

[2] J.F. Scott, Journal of Phys: cond matt., **18**, R361 (2006).

[3] B.Nagaraj, S.Aggarwal, T.K. Song, T.Sawhney and R. Ramesh, Phys. Rev. B, **59**, 16022 (1999).

[4] P.Zubko, D.J. Jung and J.F. Scott, J. Appl. Phys., **100**, 114113 (2006).

[5] H. Hu and S.B.Krupanidhi, J. Mater. Res., **9**, 1014 (1994).

[6] J.G.Simmons, Phys. Rev. Lett., **15**, 967 (1965).

[7] J.Wang, J.B.Neaton, H.Zheng, V.Nagarajan, S .B. Ogale, B.Liu, D.Viehland, V.Vaithyanathan, D.G.Schlom, U.V.Waghmare, N.A.Spaldin, K.M.Rabe, M.Wuttig and R.Ramesh, Science, **299**, 1719 (2003).

[8] W.Prellier, M.P. Singh and P. Murugavel, Journal of Physics: cond. Matt., **17**, R803 (2005).

[9] G.W. Pabst, L.W. Martin, Y.H. Chu and R.Ramesh, Appl. Phys. Lett., **90**, 072902 (2007).

[10] X.Qi, J. Dho, R.Tomov, M.G.Blamire and J.L.MacManus-Driscoll, Appl. Phys. Lett., **86**, 062903 (2005).

[11] M.A. Khan, T.P. Comyn and A.J. Bell, Appl. Phys. Lett., **92**, 072908 (2008).

[12] R.Ranjith, B.Kundys and W. Prellier, Appl. Phys. Lett., **91**, 222904 (2007),

[13] S. Bose and S.B. Krupanidhi, Appl. Phys. Lett., **90**, 212902 (2007).

[14] S.Zafar, R.E.Jones, Bo Jiang, B.White, V.Kaushik and S.Gillespie, Appl. Phys. Lett., **73**, 3533 (1998).

[15] K. Takahashi, N. Kida and M. Tonouchi, Phys. Rev. Lett., **96**, 117402 (2006).

[16] K.Van Benthem, C.Elsasser and R.H. French, J. Appl. Phys., **90**, 6156 (2001).

[17] M.A. Lampert, Phys. Rev, **103**, 1648 (1956).
8

**Figure Captions**

Figure 1. (color online) Current–Voltage characteristics of a $(BiFeO_3)_m(SrTiO_3)_m$ superlattice structure with a periodicity of $\Lambda= 40$Å, m = 5. Figure inset shows the room temperature I-V curve on reversal of the bias.

Figure 2. (color online) Modified Schottky plot of $(BiFeO_3)_m(SrTiO_3)_m$ superlattice structures with m = 5 and 7 at 323 K. Figure inset shows the activation plot of m=5 superlattice structure derived from the modified Schottky equation at different applied electric field. (Solid lines show the fitting in both the main figure and the figure inset)

Figure 3. (color online) Poole Frenkel emission plot of $(BiFeO_3)_m(SrTiO_3)_m$ superlattice structures with m = 5 at different temperatures. Figure inset shows the Poole Frenkel activation plot of a m = 5 superlattice structure. (Solid line shows the fitting)

Figure 4. (color online) High temperature Log(J)-Log(V) plot of a $(BiFeO_3)_m(SrTiO_3)_m$ superlattice structure with m = 5, exhibiting a power law dependence. (Solid line shows the fitting)